\documentclass[prb,twocolumn,amsmath,showpacs]{revtex4}

\usepackage{epsfig}
\usepackage{graphicx}

\begin{document}
\input epsf

\title {Comment on "Hole concentration dependence of penetration depth and upper critical field in Bi$_2$Sr$_2$(Ca,Y)Cu$_2$O$_{8+\delta}$ extracted from reversible magnetization"}

\author {I. L. Landau$^{1,2}$ H. Keller$^{1}$}
\affiliation{$^{1}$Physik-Institut der Universit\"at Z\"urich, Winterthurerstrasse 190, CH-8057 Z\"urich, Switzerland}
\affiliation{$^{2}$Kapitza Institute for Physical Problems, 117334 
Moscow, Russia}

\date{\today}

\begin{abstract}

We argue that the method, which was used for the analysis of equilibrium magnetization data in the original publication, is not adequate to the experimental situation. As a result, the temperature dependencies of the upper critical field $H_{c2}(T)$ and the magnetic field penetration depth $\lambda (T)$, obtained in this work, are incorrect. Using a different approach, we reanalyze the presented experimental data and demonstrate that the normalized $H_{c2}(T)$ curves are rather different from those presented in the original publication and do not follow predictions of the Werthamer-Helfand-Hohenberg theory. Another interesting observation is that the $H_{c2}(T)$ dependencies for two samples with close levels of doping are rather different.

\end{abstract}

\maketitle

In Ref. \onlinecite{kimcheon}, the equilibrium magnetization $M(H,T)$ in the mixed state of several polycrystalline Bi$_2$Sr$_2$(Ca,Y)Cu$_2$O$_{8+\delta}$ (Bi-2212) samples was investigated. The results were analyzed by employing the vortex fluctuation model.\cite{blk,kogan,kogan2} We disagree with this analysis. Below, we reanalyze the data of Ref. \onlinecite{kimcheon} by a more appropriate approach and demonstrate that temperature dependencies of the upper critical field $H_{c2}$ are rather different from those obtained in the original publication. This means that the conclusion of Ref. \onlinecite{kimcheon} that $H_{c2}(T=0)$ has a maximum value when the hole concentration $p$ is around 0.19 is not really based on presented experimental data.

There are several theoretical models,\cite{blk,hclem, brandt}  which are usually employed for evaluation of $H_{c2}$ from experimental magnetization data. All these models assume conventional superconductivity (an isotropic superconducting order parameter) and a uniform sample with a zero demagnetizing factor $n$. Neither of these conditions is satisfied in polycrystalline HTSC's. Because the differences between theoretical assumptions and experimental situations are rarely discussed in the literature, we consider them in some details. 

(i) {\it Demagnetizing factor.} 
If the sample magnetization $M$ is much smaller than an applied magnetic field $H$, demagnetizing effects are usually neglected. This is not correct. The relation between $H$ and the magnetic induction $B$ in the sample is always dependent on the value of $n$. Thus, if $n = 0$, $B \equiv H$ and $M \equiv 0$, independent of the applied magnetic field, temperature or the nature of the sample.

(ii) {\it Pairing symmetry.}
Symmetry of the order parameter is also important. In the case of unconventional $d$-pairing, which is expected in high-$T_c$ superconductors (HTSC), the distribution of the order parameter around vortex cores and the corresponding contribution to the free energy is different from that for conventional superconductors. This is why theoretical calculations based on conventional $s$-pairing should be used  with caution if they are applied for the analysis of experimental data collected on unconventional superconductors

(iii) {\it Polycrystalline samples.}
HTSC's are highly anisotropic. In such materials, if the direction of an external magnetic field does not coincide with one of the principle axes of the crystal, the magnetic induction in the sample is not exactly parallel to the applied magnetic field. In samples consisting of randomly oriented grains, this leads to an additional free energy and may influence the sample magnetization. It should also be noted that, because magnetizations of different grains are different, there is some magnetic interaction between the neighboring grains. The situation is even more complex  at higher temperatures.  Indeed, according to calculations of Brandt,\cite{brandt} the magnetic field dependence of $M$ is a linear function of $\ln H$ (London limit) only in magnetic fields $H < 0.1H_{c2}$ (see also Fig. 3 in Ref. \onlinecite{lo-comm}).  At temperatures, $T \gtrsim 0.8T_c$, the upper limit of the magnetic field range is usually higher than this value. In this case, deviations of $M(H)$ from the predictions of the London model have to be accounted for and a simple averaging, which was proposed in Ref.  \onlinecite{kogan} and used in Ref. \onlinecite{kimcheon}, is not applicable.

In order to avoid above-mentioned uncertainties, we use here a different method of analyzing magnetization data. In this scaling approach, developed in Ref. \onlinecite{lo1},  no particular $M(H)$ dependence is assumed {\it a priori}, and it can be applied to single crystals as well as to polycrystalline samples, independent of the pairing mechanism or the sample geometry.\cite{lo-c} The disadvantage of this analysis is that it does not provide the absolute values of $H_{c2}$ but only its relative variations. This is the prize to pay for its universality.  

The scaling procedure is based on the assumption that the Ginzburg-Landau parameter $\kappa$ is temperature independent.\cite{fn1} In this case, the mixed-state magnetic susceptibility may be written as
\begin{equation}
\chi(H,T) = \chi(H/H_{c2}),
\end{equation}
i.e., the temperature dependence of  $\chi$ is only due to temperature variation of $H_{c2}$. Eq. (1) is already sufficient to establish a relation between magnetizations at two different temperatures\cite{lo1} 
\begin{equation}
M(H/h_{c2},T_0) = M(H,T)/h_{c2}, 
\end{equation}
where $h_{c2}(T) = H_{c2}(T)/H_{c2}(T_0)$ is the upper critical field normalized by its value at $T = T_0$. This equation is valid if the diamagnetic response of the mixed state is the only significant contribution to the sample magnetization. Considering HTSC's, however, we have to take into account their noticeable paramagnetic susceptibility $\chi_n$ in the normal state and its dependence on temperature. In order to account for $\chi_n(T)$, we have to introduce an additional $c_0(T)H$ term in Eq. (2). According to Ref. \onlinecite{lo1}, the resulting equation connecting $M(H,T_0)$ and $M(H,T)$ may be written as
\begin{equation}
M(H/h_{c2},T_0) = M(H,T)/h_{c2} + c_0(T)H
\end{equation}
with
\begin{equation}
c_0(T) = \chi_n(T) - \chi_n(T_0)
\end{equation}

We note that Eqs. (1) and (2) are rather general and they can be obtained from any model based on the Ginzburg-Landau theory, including the so-called nonlocal London theory and the Hao-Clem model. At the same time, as was discussed in Refs. \onlinecite{lo-hcl} and  \onlinecite{lo-lt2}, these relations remain valid even if the $M(H)$ dependence is different from predictions of the Ginzburg-Landau theory for conventional superconductors.

Eq. (3) can be used as the basis for the scaling procedure. The adjustable parameters $h_{c2}(T)$ and $c_0(T)$ are obtained from the condition that $M(H/h_{c2},T_0)$, calculated from data collected at different temperatures, collapse onto a single master curve, which represents the equilibrium magnetization at $T = T_0$.\cite{lo1} It was demonstrated that this scaling procedure works quite well and may be used in order to obtain temperature dependencies of the normalized upper critical field $h_{c2}$ from equilibrium magnetizations measured at different temperatures.\cite{lo1,lo-c,lo-hcl,lo-comm,lo-lt2,thomp} In the following, we use $M_{eff}(H)$ to denote $M(H/h_{c2},T_0)$ calculated by using Eq. (3) in order to distinguish it from directly measured magnetization data.

\begin{figure}[h]
 \begin{center}
  \epsfxsize=0.95\columnwidth \epsfbox {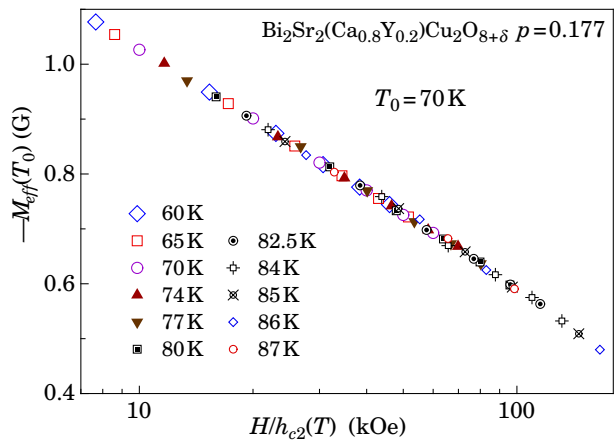}
  \caption{The scaling results for a Bi$_2$Sr$_2$(Ca$_{0.8}$Y$_{0.2}$)Cu$_2$O$_{8+\delta}$ sample. Magnetization data are taken from Fig. 1 of Ref. \onlinecite{kimcheon}.}
 \end{center}
\end{figure}
Figs. 1 and 2 show the scaled magnetization curves for two Bi$_2$Sr$_2$(Ca$_x$,Y$_{1-x}$)Cu$_2$O$_{8+\delta}$ samples with different hole concentrations $p$. Magnetization data are taken from Ref. \onlinecite{kimcheon}. The quality of scaling is almost perfect in both cases and deviations of individual data points do not exceed the accuracy of the original data as they can be taken from the figures presented in Ref. \onlinecite{kimcheon}.
\begin{figure}[h]
 \begin{center}
  \epsfxsize=0.95\columnwidth \epsfbox {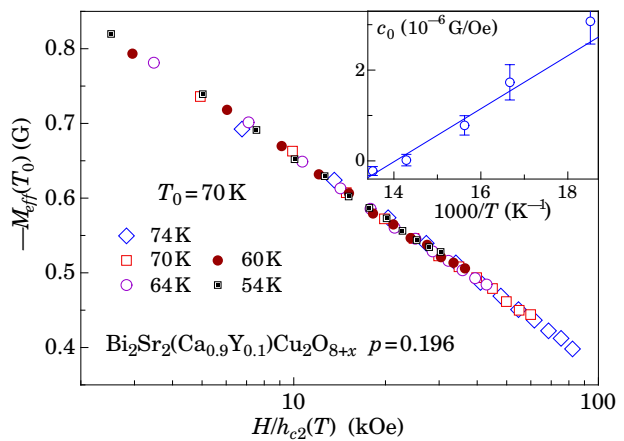}
  \caption{The scaling results for a Bi$_2$Sr$_2$(Ca$_{0.9}$Y$_{0.1}$)Cu$_2$O$_{8+\delta}$ sample. Magnetization data are taken from Fig. 2 of Ref. \onlinecite{kimcheon}. The inset shows the scaling parameter $c_0$ versus $1/T$. The straight line is the best linear fit to data points. Error bars are estimated by the fitting procedure and do not include possible experimental errors.}
 \end{center}
\end{figure}

The temperature dependence of the scaling parameter $c_0$ for one of the samples is plotted in the inset of Fig. 2  as a function of $1/T$. As expected, $c_0$ is small and its dependence is close to the Curie-Weiss law. Because $c_0$ describes only a small correction to the sample magnetization, uncertainty of this parameter is rather high. 
\begin{figure}[h]
 \begin{center}
  \epsfxsize=0.95\columnwidth \epsfbox {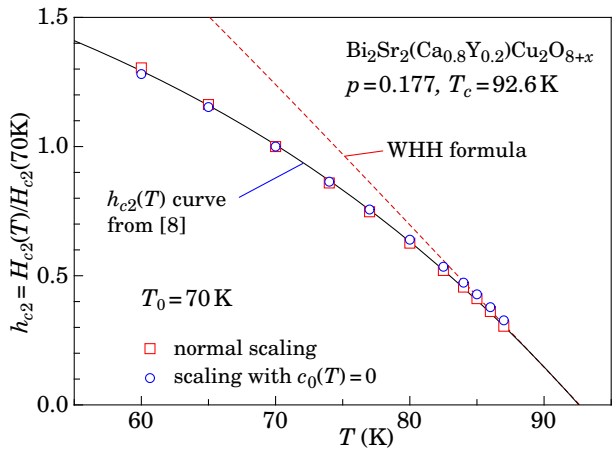}
  \caption{The normalized upper critical field $h_{c2}$ as a function of temperature  for a Bi$_2$Sr$_2$(Ca$_{0.8}$Y$_{0.2}$)Cu$_2$O$_{8+\delta}$ sample. For comparison, we also show $h_{c2}(T)$ obtained if the scaling is based on Eq. (2) ($c_0(T)  \equiv 0$). The solid line represents the best fit of the "universal" $h_{c2}(T/T_c)$ curve, obtained in Ref. \onlinecite{lo1}, to the data points (see text for details). The dashed line corresponds to the WHH theory.}
 \end{center}
\end{figure}

The resulting temperature dependencies of the normalized upper critical field for these samples are shown in Figs. 3 and 4. We note that uncertainty of $h_{c2}$ is much smaller than that for $c_0$ and the corresponding error bars are smaller than the size of symbols. In order to demonstrate a weak interference between the two fit parameters, we repeated the scaling procedure assuming $c_0 \equiv 0$. As may be seen in Fig. 3, the difference between the two sets of data-points is insignificant (see also Ref. \onlinecite{thomp}). At the same time, because the normal-state paramagnetism in HTSC's exists and it is temperature dependent, we do not see any reason to neglect $c_0(T)$.  

\begin{figure}[h]
 \begin{center}
  \epsfxsize=0.95\columnwidth \epsfbox {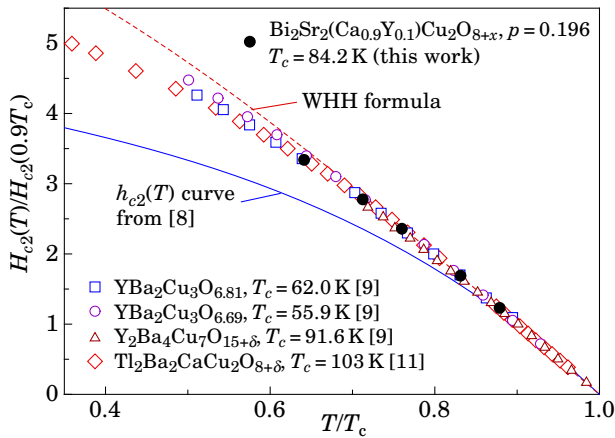}
  \caption{The normalized upper critical field $h_{c2}$ as a function of $T/T_c$  for a Bi$_2$Sr$_2$(Ca$_{0.9}$Y$_{0.1}$)Cu$_2$O$_{8+\delta}$ sample. The $h_{c2}(T/T_c)$ curves for several different samples, obtained by employing the same scaling procedure,\cite{lo-c,lo-hcl}, are shown for comparison. The solid and the dashed lines are the same as in Fig. 3.}
 \end{center}
\end{figure}
In Refs. \onlinecite{lo1} and  \onlinecite{lo-c}, it was shown that all numerous HTSC's may be divided into two groups. The $h_{c2}(T/T_c)$ curves for HTSC's belonging to the same group are practically identical, while they are distinctly different between the groups. Acknowledging the fact that  the larger group includes a huge variety of different HTSC compounds, while the second one is rather small and apparently limited to a very few particular HTSC's,\cite{lo1,lo-c,lo-hcl,lo-comm} we shall denote the corresponding $h_{c2}(T/T_c)$ curves as {\it typical} and {\it unusual}, respectively. Quite surprisingly, for the two samples analyzed in this work, the $h_{c2}(T)$ dependencies turned out to be different and, as we discuss below, these dependencies perfectly match the corresponding curves for the two above mentioned groups of HTSC's.

The results for a sample with the hole concentration $p = 0.177$ are shown in Fig. 3. The solid line represents the "typical" $h_{c2}(T/T_c)$ dependence, obtained in Ref.  \onlinecite{lo1}, and fitted to the data points by adjusting $T_c$. $T_c = 92.6$ K, evaluated in such a way, is close to the value given in the original publication. As may be seen in Fig. 3, the data points follow the solid line quite closely clearly demonstrating that this sample belongs to the larger group of HTSC's.   We also note that because the $h_{c2}(T)$ curve for this group of HTSC's is completely different from predictions of the Werthamer-Helfand-Hohenberg (WHH) theory (the dashed line in Fig. 3),\cite{wert} the zero-temperature upper critical field $H_{c2}(0)$, evaluated by employing this theory will be well above its real value.  

The $h_{c2}(T/T_c)$ curve for a sample with $p = 0.196$, as shown in Fig. 4, is quite different and practically coincides with those for several other HTSC compounds belonging to the smaller group of HTSC's. Our evaluation of $T_c = 84.2$ K for this sample practically coincides with the value provided in Ref. \onlinecite{kimcheon}.  We note that, although in this case $h_{c2}(T/T_c)$ is closer to the WHH theory, the differences are still significant and this theory should not be used for the evaluation of $H_{c2}(0)$.  

The observation that, considering temperature dependencies of $H_{c2}$, all numerous HTSC's may be divided into two groups is remarkable. However, it is even more surprising that no any intermediate $h_{c2}(T/T_c)$ was observed so far. Because both typical and unusual $h_{c2}(T/T_c)$ curves were observed in the same families of HTSC's,\cite{lo1,lo-c,lo-comm} one may assume that the level of doping is essential. A similar conclusion may also be drown from the results presented in this work. This is why it would be would be extremely interesting to study the transition from one type of the $h_{c2}(T/T_c)$ dependence to the other systematically.

The model of thermal fluctuations of vortices,\cite{blk} which was used for the analysis of experimental data in Ref. \onlinecite{kimcheon}, is based on two prominent experimental observations. (i) In the case of layered Bi- and Tl-based HTSC compounds, there is a temperature $T^* < T_c$, at which the sample magnetization does not depend on an applied magnetic field.\cite{kes} (ii) If the Hao-Clem model\cite{hclem}  is applied for analyzing of magnetization data, the resulting temperature dependence of $\kappa$ always demonstrates a strong and unphysical increase with increasing temperature.\cite{hc1,hc2,hc3,hc4,hc5,hc6,hc8,hc9,hc10,hc11,hc12,hc13,hc14} This increase is especially strong for layered HTSC's.\cite{hc13,hc14} Both these features were interpreted as evidences that thermal fluctuations in the mixed-state of layered HTSC's remain strong down to temperatures well below $T_c$.\cite{blk} 

There is no doubt that fluctuation effects are strong at temperature close to $T_c$ and it is possible that they are responsible for the crossing of the $M(T)$ curves at $T = T^*$. As well as we are aware, the BLK model is the only theory describing this feature. Using this approach, one may evaluate the magnetic field penetration depth $\lambda$ at $T = T^*$. The only parameter $s$, entering the expression for $\lambda(T^*)$, represents the distance between superconducting layers. This parameter may be independently evaluated as $s = - (k_B T^*)/(\Phi _0 M^*)$ ($k_B$ is the Boltzmann constant, and $\Phi _0$ is the magnetic flux quantum). However, the ratio $-T^*/M^*$ is always smaller than the theoretically predicted value and, contrary to the theory, $T^*/M^*$ is practically independent of $s$.\cite{xue} This is why the BLK model may be considered only as a qualitative approach to the problem and the value of $\lambda(T^*)$ resulting from this model may be different from the actual magnetic field penetration depth.

Although the fluctuation approach is often used for the analysis of experimental magnetization data even at temperatures well below $T^*$, there are no independent evidences of its applicability in this temperature range. The main argument in favor of strong fluctuation effects is a breakdown of the Hao-Clem model. However, as was discussed in Refs. \onlinecite{lo-hcl} and \onlinecite{lo-comm}, one can find a number of other reasons in order to explain this result. Furthermore, alternative analysis of experimental data, presented in this work, clearly demonstrates that fluctuations induced corrections to the sample magnetization are only observable in a rather narrow temperature range below $T^*$.

In conclusion, using an alternative approach to the analysis of experimental data presented in Ref. \onlinecite{kimcheon}, we demonstrate that the temperature dependencies of the upper critical field are rather different from  the results of the original publication. The same is applied to the temperature dependencies of the magnetic field penetration depth. This means that the dependence of $H_{c2}(0)$ on the doping level obtained in  Ref. \onlinecite{kimcheon} cannot be considered as based on the experimental results. It is also interesting that two Bi-2212 samples with close doping levels exhibit rather different temperature dependencies of $H_{c2}$.

This work is partly supported by the Swiss National Science Foundation.


\begin{thebibliography}{71}

\bibitem{kimcheon} G. C. Kim, M. Cheon, H. Kim, Y.C. Kim, D. Y. Jeong , Phys. Rev. B {\bf 72}, 64525 (2005).

\bibitem{blk} Z. Te\v{s}anovi\'c, L. Xing, L. Bulaevskii, Q. Li, M. Suenaga, Phys. Rev. Lett. {\bf 69}, 3563 (1992); L. N. Bulaevskii, M. Ledvij, V. G. Kogan, Phys. Rev. Lett. {\bf 68}, 3773 (1992).

\bibitem{kogan} V. G. Kogan, M. Ledvij, A. Yu. Simonov, J. H. Cho, D. C. Johnston, Phys. Rev. Lett. {\bf 70}, 1870 (1993).  

\bibitem{kogan2} V. G. Kogan, A. Gurevich, J. H. Cho, D. C. Johnston, M. Xu, J. R. Thompson, and A. Martynovich, Phys. Rev. B {\bf 54}, 12386 (1996).

\bibitem{hclem} Z. Hao, J. R. Clem, M. W. McElfresh, L. Civale, A. P. Malozemoff, F. Holtzberg, Phys. Rev. B {\bf 43}, 2844 (1991).

\bibitem{brandt} E. H. Brandt, Phys. Rev. Lett. {\bf 78}, 2208 (1997).  E. H. Brandt, Phys. Rev. B {\bf 68}, 054506 (2003).

\bibitem{kes} P. H. Kes, C. J. van der Beek, M. P. Maley, M. E. McHenry, D. A. Huse, M. J. V. Menken, A. A. Menovsky, Phys. Rev. Lett. {\bf 67}, 2383 (1991). 

\bibitem{lo1} I. L. Landau and H. R. Ott, Phys Rev. B {\bf 66}, 144506 (2002).

\bibitem{lo-c} I. L. Landau and H. R. Ott, Physica C {\bf 385}, 544 (2003).

\bibitem{fn1} $\kappa = \lambda(T)/\xi(t)$ where $\lambda(T)$ and $\xi(T)$ are the magnetic field penetration depth and the coherence length, respectively. In fact, the temperature independence of the ratio $\lambda(T)/\xi(T)$ is a sufficient condition for the applicability of this scaling procedure. This is why it can be used even outside the temperature range of quantitative applicability of the Ginzburg-Landau theory.

\bibitem{lo-hcl} I. L. Landau and H. R. Ott, Physica C {\bf 411}, 83 (2004).

\bibitem{lo-comm} I. L. Landau and H. R. Ott, Phys Rev. B {\bf 72}, 176502 (2005).

\bibitem{lo-lt2} I. L. Landau and H. R. Ott, J. of Low Temp. Phys. {\bf 139}, 175 (2005).

\bibitem{thomp} J. R. Thompson, J. G. Ossandon, L. Krusin-Elbaum, D. K. Christen, H. J. Kim, K. J. Song, K. D. Sorge, and J. L. Ullmann, Phys. Rev. B {\bf 69}, 104520 (2004).

\bibitem{wert} E. Helfand and N. R. Werthamer, Phys Rev. {\bf 147}, 288 (1966), N. R. Werthamer, E. Helfand and G. Hohenberg, {\it ibid.} {\bf 147}, 295 (1966).

\bibitem{xue} Y. Y. Xue, Y. Cao, Q. Xiong, F. Chen, C. W. Chu, Phys. Rev. B {\bf 53}, 2815 (1996).

\bibitem{hc1} Q. Li, M. Suenaga, J. Gohng, D. K. Finnemore, T. Hikata, K. Sato, Phys. Rev. B {\bf 46}, R3195 (1992).

\bibitem{hc2} Qiang Li, M. Suenaga, T. Kimura, and K.Kishio, Phys. Rev. B {\bf 47}, 2854 (1993).

\bibitem{hc3} D. N. Zheng, A. M. Campbell, and R. S. Liu, Phys. Rev. B {\bf 48}, 6519 (1993).

\bibitem{hc4} Y. C. Kim,  J. R. Thompson, J. G. Ossandon, D. K. Christen, M. Paranthaman, Phys. Rev. B {\bf 51}, 11767 (1995).

\bibitem{hc5} Mung-Seog Kim, Sung-Ik Lee, Seong-Cho Yu, and Nam H. Hur,  Phys. Rev. B {\bf 53}, 9460 (1996).

\bibitem{hc6} J. R. Thompson, J. G. Ossandon, D. K. Christen, M. Paranthaman, E. D. Specht, and Y. C. Kim, Phys. Rev. B {\bf 54}, 7505 (1996).

\bibitem{hc8} Y. Zhuo, J. H. Choi, M. S. Kim, W. S. Kim,  Z. S. Lim, S. I. Lee, S. Lee, Phys. Rev. B {\bf 55}, 12719 (1997).

\bibitem{hc9} Mung-Seog Kim, Sung-Ik Lee, Seong-Cho Yu, I. Kuzemskaya, E. S. Itskevich, and K. A. Lokshin,  Phys. Rev. B {\bf 57}, 6121 (1996).

\bibitem{hc10} M. Y. Cheon, G. C. Kim, B. J. Kim, and Y. C. Kim, Physica C {\bf 302}, 215 (1998).

\bibitem{hc11} Yi Zhuo, Su-Mi Oh, Jae-Hyuk Choi, Mun-Seog Kim, Sung-Ik Lee, N. P. Kiryakov, M. S. Kuznetsov, and Sergey Lee, Phys. Rev. B {\bf 60}, 13094 (1999).

\bibitem{hc12} Heon-Jung Kim, P. Chowdhury, In-Sun Jo, and Sung-Ik Lee, Phys. Rev. B {\bf 66}, 134508 (2002).

\bibitem{hc13} J. H. Cho, Zhidong Hao, and D. C. Johnston, Phys. Rev. B {\bf 46}, R8679 (1992).

\bibitem{hc14} K.-H. Kim, H.-J. Kim, S.-I. Lee, A. Iyo, Y. Tanaka, K. Tokiwa, T. Watanabe, Phys. Rev. B {\bf 70}, 92501 (2004).

\end{thebibliography}
\end{document}